# TRANS-Net: an Efficient Peer-to-Peer Overlay Network Based on a Full Transposition Graph

Stavros Kontopoulos
Comp. Eng. & Inf. Dept.
University of Patras
kontopou@ceid.upatras.gr

Athanasios Tsakalidis
Comp. Eng. & Inf. Dept.
University of Patras
& Computer Technology Institute
Patra, PO 22500
tsak@cti.gr

## Abstract

*In this paper we propose a new practical P2P system based on a full transposition network topology named TRANS-Net. Full transposition networks achieve higher fault-tolerance and lower congestion among the class of transposition networks. TRANS-Net provides an efficient lookup service i.e. $k$ hops with high probability where $k$ satisfies $\Theta(\log_n m) < k < \Theta(\log_2 m)$, where m denotes the number of system nodes and n is a system parameter related to the maximum number that m can take (up to n!). Experiments show that the look-up performance achieves the lower limit of the complexity relation. TRANS-Net also preserves data locality and provides efficient look-up performance for complex queries such as multi-dimensional queries.*

## 1. Introduction and related work

P2P systems emerged as a class of distributed systems focusing on load balancing and providing guarantees for the lookup process counted in number of hops in the virtual topology. In our work we introduce a new protocol that uses a transposition network with nodes arranged also in a ring topology. Our system keeps the order of data and supports one dimensional range queries, multi-dimensional queries. Most structured P2P systems fall under the category of distributed hash tables (DHTs) which form an overlay network that maps keys to network nodes and objects using some consistent hashing function. Some popular DHTs are CAN[11], Chord[14], Pastry[12], Tapestry[1], Koorde[4]. Some P2P systems fall under the category of tree based systems. The latest, Baton* [3] is a tree based system that exhibits $O(\log_t m)$ lookup performance with O(*t*) routing space overhead where *t* is the tree fanout. Baton* is a non-trivial improvement over the Baton [2] system with Baton being a similar tree based system with fan-out 2. All results are achieved at the presence of *m* nodes in the system. The query support of our system is not available to DHTs systems. Compared to Baton* our system is based on a symmetrical topology which guarantees that traffic will be equally shared among available links in a flash crowd scenario. Moreover our system is more fault-tolerant than Baton*. Koorde is a P2P system based on a sparse De Bruijn network embedded in a ring. The sparse network is a necessary condition for Koorde to guarantee optimal performance. Compared to Koorde our system works efficiently for both dense and sparse networks. There are three basic topologies based on transpositions: star networks [8], [9], pancake networks [8], [9] and full transposition networks [6]. Pancake networks were used as overlay networks in [5], [10]. Compared to [5], our system is optimized from a more practical perspective avoiding the usage of moving virtual peers and data order is preserved. Moreover in [10] a P2P system is proposed that uses a tree based framework to achieve O(1) load balancing w.h.p. The framework emulates a pancake overlay. Again here virtual nodes are necessary. An interesting issue here is whether full transposition networks can be emulated by a similar and/or other frameworks. The rest of the paper is structured as follows. In section 2 we define full transposition networks. In sections 3 to 8 we introduce our new P2P system design based on full transposition networks and discuss parameters of our design such as basic system operations, load balancing, multi-dimensional query support, load-balancing, fault-tolerance and congestion. In section 9 we analyze the performance of the new overlay network. Section 10 concludes the paper and describes future work.

Overall, our *contribution* is that:

- We introduce a new practical, competitive yet simple P2P system based on an innovative design, using a transposition network.
- Our system simultaneously optimizes or achieves good values for several P2P design parameters leading to a good overall design compared to other systems.
- We show how our system explicitly supports multi-dimensional queries, without proper modification or extension.

Since our system does not fall to a specific category we have chosen to compare it with the popular P2P system Chord when that is applicable.

## 2. Preliminaries

### 2.1. Notation

In this context with $m$ we denote the number of peers in a P2P system, $n$ is a system parameter that is used to represent the number of symbols that constitute a permutation. As it is described in the following sections, $n$ influences the number of peers in a static star network which is $n!$ and inevitably the maximum number of peers, again $n!$, participating in our proposed P2P system, TRANS-Net.

### 2.3. Full Transposition Networks

**Definition 1** Let $G$ be the finite group of permutations on a set with $n$ elements numbered as $\{1, 2, \ldots n\}$ and $S$ be a set of generators for the group closed under inverses with $S$ being:

$S = \{\pi_{i,j}(\cdot) \mid p \in P, \pi_{i,j}(p) \in P\}$ with

$\pi_{i,j}(p) = p_1 p_2 \cdots p_{j-1} p_i p_{j+1} p_{i-1} p_j p_{i+1} \cdots p_n$,

$p = p_1 p_2 \cdots p_j p_i \cdots p_n$

where P is the set of all possible permutations.

Then the Full Transposition Network (FTN) on $G$ is the graph with vertex set $G$ and edge set $E$ defined as follows $E = \{(p, \pi_{i,j}(p))\}, \forall i, j, \in \{1, 2, .., n\}, i \neq j$.

The number of the unique links is $\frac{n^2 - n}{2}$. Basic properties of this graph, which is also a Cayley graph, such as diameter and fault-tolerance are analyzed here [6]. FTN has a diameter of n-1. This small diameter is what we seek to exploit in our routing algorithm. In order to use a FTN network for one dimensional exact query matching we must provide a routing algorithm which finds a path from a certain node to another node. Routing algorithms on transposition networks, usually exploit the underlying mathematical structure of permutations which leads to optimal routing design as in [15]. In our approach a greedy algorithm is chosen for fault-tolerance issues and its simplicity and immediate applicability in a P2P system. On a FTN network with $n!$ nodes we can route a message from a node $X = x_1 \ldots x_n$ to some node $Y = y_1 \ldots y_n$ by means of transpositions at each step correcting a digit of $X$ from left to right until $X$ is replaced by $Y$. Since all transpositions are available we can route a message in $O(n)$ steps.

## 3. A Peer-to-Peer System Based on FTNs.

FTN networks are static networks. In the field of P2P system design it is necessary for the topology of the system to be dynamic. In other words it cannot be assumed that all peers will be present in the system. In order to cope with that we simply embed a FTN network in a Chord like ring. When a node links to some missing node it replaces it with its immediate successor on the ring. This way we ensure that we can approach greedily the target at the absence of some nodes in the network transforming the system to a dynamic one. Now we are ready to describe a P2P system based on the modified FTN network just described. We call this modified FTN network TRANS-Net.

### 3.1. System model

Let $n$ be a parameter of the system. Each peer has a unique identifier taken from the set of $n!$ permutations using some hashing function and then the permutation identifier is mapped to a number in the key space $[N_{left}, N_{right}]$. The choice of the map function and of the appropriate values of $N_{left}, N_{right}$ will be discussed in section 5 (the load balancing section). Collisions here can be checked by a simple lookup of the key generated by the hashing function or they can be avoided with the proper value of $n$. Data elements in the network are indexed by an identifier taken from the interval $[N_{left}, N_{right}]$. Each peer $v$ is responsible for the keys that belong to the interval $(v.predecessor, v]$ where v.predecessor is the closest preceding node to *v*. Our system inherits the good properties of the Chord ring like dynamic manipulation of node joins and leaves.

## 4. System Operations

TRANS-Net offers the basic operations found in most DHTs systems which are key lookup, key insertion/deletion and peer join/leave. These operations are described here. The key lookup is performed by locating the successor peer on the ring for a key following FTN links (star links). Here we introduce two algorithms for searching in a TRANS-Net.

### 4.1. Greedy Algorithm

The first routing algorithm finds among the available links of the current node the one that leads us closer to the target node. This is done by checking its routing table. Comparisons for pickng the appropriate link are made using the virtual distance from the target node and specifically from the left and right bounds of the key space allocated to the candidate node for the next step. This is necessary for the correctness of our algorithm and it implies that we hold for each routing table entry, additional information for the key space managed by the node in that entry. The pseudocode of the greedy algorithm is given next:

---
**Algorithm 1** Greedy_Search_key (x, k)

**Input**: x is current node, k is the key to be found

**Output:** node that holds k

//O is the node that owns key k if the network had n! //nodes.

1. Node O=owner (k);
2. Node C=find_closest_to_target_in_route_table(O);
3. Greedy_Search_key (C, k);

**Figure 1. TRANS-Net's greedy algorithm.**

---

The search algorithm ensures that the whole procedure converges to the target peer.

### 4.2. Heuristic Algorithm

Along with the above algorithm we implemented a heuristic that corrects one digit of the current node's permutation id according to the target node's id. This algorithm may fail to track the target node but this fact does not make the algorithm useless since its hit rate is high according to our experimental results (see experimental section). The following theorems indicate the performance of both algorithms. In a TRANS-Net network with *m* peers routing a message is accomplished in $O(n)$ hops in the worst case. In the average case the routing algorithm performs much better depending only on the current number *m* of peers. We have the following theorem:

**Theorem 1:** In a TRANS-Net network with *m* peers routing a message is accomplished in $k$ hops in the average case where k satisfies $\Theta(\log_n m) < k < \Theta(\log_2 m)$.

**Proof.** The set of all possible permutations with each permutation having *n* symbols can be distributed to *n* classes according to the leftmost symbol. Each class can be further divided to subclasses according to the next digit from the left. The *i*-th digit from the left divides the previous class to *n-i* subclasses. We proceed until we have subclasses of size 1. Let *m* be the number of peers in the system. Now suppose we choose for those *m* peers randomly a permutation identifier from the set of *n*! possible permutations. On the average case the permutation identifiers are evenly distributed to all classes. For example each member of the first group of classes according to the leftmost symbol should have on the average $\frac{m}{n}$ peers. Let *X* be a peer who searches for a key *q* which is owned by peer *Y*. We must mention here that the permutation id of the node *T* which would have owned q if n! nodes were present belongs to *Y*'s key space too. Let *X* be the first peer on the circle and *Y* be the last one. Hence, we have the maximum possible distance to the target which is $D_0 = m$ in terms of the number of peers. In the first step of the routing algorithm the leftmost digit is corrected, we get closer to the target and the distance becomes: $D_1 = m - \frac{m}{n}(n-1) = \frac{m}{n}$. In the second step the distance becomes $D_1 = \frac{m}{n} - \frac{m}{n(n-1)}(n-2) = \frac{m}{n(n-1)}$. In the final *k*-th step we have for the distance: $D_k = \frac{m}{n(n-1)(n-k+1)} \leq 1$.

It holds that: $\frac{m}{n^k} < \frac{m}{n(n-1)(n-k+1)} \leq \frac{m}{2^k}, n > 3$ which implies that $\log_n m < k \leq \log_2 m$. This analysis is enough for the heuristic algorithm. In order to extend to the greedy algorithm we need to take into consideration the fact that it is possible that the closest distance may not be the one that leads to a node that corrects the appropriate digit according to T's id. This may lead to a slow-down of the algorithm as a matter of fact it may lead to a serialized movement on the ring. We clarify this by an example and show that this event is rare and hopefully it can be tracked and avoided. Suppose the initial node for the key look up is $X$=365241 for a TRANS-Network with n=6 and we search for a key for which $T$'s id is 413562 (see next figure).

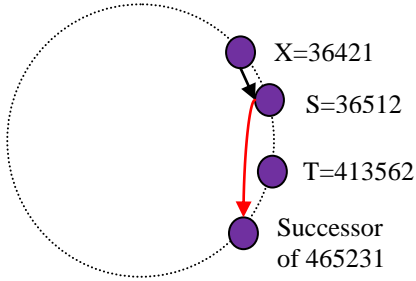

**Figure 2. Example for proof of theorem 2.**

It is obvious that our algorithm will choose the successor of node 365241 which is $S$=365124 as closest to the target (fig.2 black arrow), instead of the successor node in its routing table of node 465231. Following such a node doesn't lead to a new node where some of its digits in its permutation id are equal to the respective digits in taget's node id. Luckily that's not a problem. As we move closer to the target inevitably we will get to a node where its digits are in accordance with those of the target. In practice, in order to move fast to a node which finally corrects the relative digit and avoid a serial move on the ring, we must detect this event. Notice here that between $S$ and $T$ there can be a large number of nodes which we want to leave behind. Detection is easy because we will be forced, according to the greedy algorithm, to move repeatedly to nodes which leave the appropriate digit incorrect and slowly lead to the target. In any of these steps we can jump to a node which corrects the digit. In the example this means that we should move to node 465231 (fig. 2, red arrow) and continue from there the key look-up. This is an adequate solution, since as we show next the event is as a matter of fact rare. Let's see why. The left most digit of the target's id should be one greater than the leftmost digit of the initial node's id. This happens with probability $\frac{1}{n}$. Moreover, the second leftmost digits of the two identifiers should have the appropriate difference counted in number of positions in the sorted permutation I, 123456 in our example. In the example the second leftmost permutation digit of the initial node is 6 while the second leftmost permutation digit of the target id is 1. The difference here is 1. This makes node's 365241 successor closer to the target than the successor of node 465231. We count the probability of having the appropriate distance for all relevant combinations of digits being at most 1 minus the probability of not having it for the cases where this is determined only from the current leftmost digit. The last probability is: $\frac{n-\frac{n}{2}}{n} = \frac{1}{2}$. Hence the total probability of the event described in the above example is at most $\frac{1}{2 \cdot n}$. Hence with high probability as n gets big enough the event is rare. This fact concludes our proof and suggests that the constant in the expression of the theorem is small as it is derived from the experimental results. Though our analysis is not tight it is obvious that the quantity $\frac{m}{n(n-1)(n-k+1)}$ is close to $\frac{m}{n^k}$ and this fact is proven from the experimental results that achieve the lower bound of the complexity of the theorem (see experimental evaluation section for more details).

**Remark**: Exploiting the order of the identifiers on the ring, the range queries of the form $[k_l, k_r]$ require $O(n+|A|)$ hops in the worst case and $O(k+|A|)$ in the average, where $|A|$ is the number of permutation-based identifiers between the peers responsible for $k_l, k_r$ respectively.

Now we will extend theorem 2 to hold with high probability.

**Theorem 2:** In a TRANS-Net network with $m$ peers routing a message is accomplished in $k$ hops with high probability where $k$ satisfies $\Theta(\log_n m) < k < \Theta(\log_2 m)$.

**Proof.** Again we use the idea of the distribution of permutations to classes according to the proof of theorem 2. Permutations are distributed uniformly at random. Let $m$ be the number of peers in the system. Initially, for the leftmost symbol of a permutation we have n classes which get $\frac{m}{n}$ permutations each, on the average. Now for $m \gg n$ it is a classical result that the maximum load will be with high probability $\Theta(\frac{m}{n})$. Within of each of these n classes the load $\Theta(\frac{m}{n})$ is distributed to n-1 classes again according to the next symbol from left to right and so on. For each distribution it holds the same bound for the maximum load except for the last few distributions for which the load is almost equal to the number of available classes. For the last case the maximum load deviation is $\Theta(\frac{\log v}{\log \log v})$ w.h.p. where v is the load. Let X be a peer who searches for a key *q* which is owned by peer Y. We have the maximum possible distance to the target which is $D_0 = m$ in terms of the number of peers. In the first step of the routing algorithm the leftmost digit is corrected, we get closer to the target and the distance becomes: $D_1 = \Theta(\frac{m}{n})$ w.h.p. in the worst case. In the second step the distance becomes $D_1 = \Theta(\frac{m}{n(n-1)})$ w.h.p in the worst case. In some step j it holds $\frac{m}{n(n-1)...(n-j+1)} \approx n$ and so the distance left is at most $\Theta(n \cdot (1 + \frac{\log n}{\log \log n}))$. In two steps from that step on we reach the destination.

We have for the total distance covered: $D_k = \Theta(\frac{m}{n(n-1)(n-k+1)}) \le 1$. It holds that: $\frac{m}{n^k} < \frac{m}{n(n-1)(n-k+1)} \le \frac{m}{2^k}$, $n > 3$ which implies that $\Theta(\log_n m) < k \le \Theta(\log_2 m)$.

### 4.3. Key Insertion/Deletion, Node join/leave

Insertion and deletion of keys is easily done by means of the above searching algorithm and it is trivial to implement them. Hence, the cost in the number of hops for insertion or deletion of a key is $O(n)$.

Each new node in order to join the system must first generate a new permutation to acquire its identifier. Then it contacts a peer which is known to be almost surely active. The peer contacted by the new peer conducts a key lookup for the identifier of the new node. Assuming no collisions, it finds the successor and informs the new node about its position on the ring. The new node enters the ring in a way similar to that in Chord by setting his predecessor and successor links and notifying for his presence both his predecessor and successor peers. After that the new peer needs to set his FTN links. In order to do this it searches for all nodes which are closest successors or predecessors to the keys produced by transpositioning the digits of its identifier relevant to some specific digit. The cost for joining the network is $O(n^3)$ in the worst case and $O(n^2 \cdot k)$ in the average. The procedure of leaving the network is simple. The peer leaving the network contacts its neighbors and informs them that it is about to leave. Since a node points to $O(n^2)$ nodes the total cost of this operation is $O(n^2)$.

### 5. Load Balancing

In our system object identifiers are chosen from the key space and the system does not impose any restriction on this procedure. Hence, the order of the data can be preserved. System nodes identifiers are chosen as follows. A new node *q*, selects in a uniformly random way, a permutation from the set of the *n*! possible permutations. Let this selected permutation be $D = [d_1, d_2, ..., d_n]$. Then this permutation is mapped to a number in the key space $[N_{left}, N_{right}]$. Here we describe two key assignment schemes in the sense that they equally distribute the load of keys to each peer. Other mapping functions are possible.

### 5.1. Key Assignment Scheme A

The first key assignment scheme represents each permutation as a *n* digit number in the *n+1*-ary number system. As a result the identifier of the new node *q* would be: $f(q) = d_1 \times {n+1}^{n-1} + d_2 \times {n+1}^{n-2} + ... + d_n \times 1$.

For this scheme, the bounds $N_{left}, N_{right}$ are set to values $0$ and ${n+1}^n - 1$ respectively. The key assignment f is a bijection and in order to use it along with the greedy algorithm described previously we implemented the reverse function finding the node that holds a particular key, owner of the key. The mapping we use here raises the question whether it produces a balanced distribution of the node identifiers over the space referenced above. We provide a simple analysis that reveals that this is in fact a balanced and simple identifier assignment scheme followed by extensive experimental results. Let $S = \{Y_1, Y_2, ..., Y_{n!}\}$ be the set of all possible permutations sorted by their value. We select randomly and uniformly $m$ distinct permutations from $S$ and let $Q = \{q_1, q_2, ..., q_m\}$ be the set of these permutations. If we pick two successive permutations $q_i, q_{i+1}$ then it holds on the average: $q_i = Y_j$, $q_{i+1} = Y_{j+\frac{n!}{m}}$. Now let $m=n$ and $Y_j = [y_{j,1}, y_{j,2}, ..., y_{j,n}]$. Also let

$Y_A = [y_{j,1}, x_1, ..., x_n]$, $Y_B = [y_{j,1}, x_n, ..., x_1]$, where $x_1 < ... < x_n$, $x_1, ..., x_n \in {1, 2, ..., n} - y_{j,1}$ and $Y_j \in C = \{Y_A, Y_{A+1}, ..., Y_B\} \subseteq S$. The set denoted by $C$ is sorted like $S$ and contains

(n-1)! permutations.

Similary $Y_{j+\frac{n!}{m}=j+(n-1)!} \in D = \{Y_C, Y_{C+1}, ..., Y_D\} \subseteq S$, where $Y_C = [y_{j,1}+1, w_1, ..., w_n]$, $Y_D = [y_{j,1}+1, w_n, ..., w_1]$ and $w_1 < ... < w_n$, $x_1, ..., x_n \in {1, 2, ..., n} - (y_{j,1}+1)$.

Obviously $Y_j$, $Y_{j+(n-1)!}$ differ in their first digit from the left and since their total distance in the permutations order is (n-1)!, their value difference $value(Y_{j+(n-1)!}) - value(Y_j)$ is $\approx (n+1)^{n-1}$. The perfect load for a node in the network is for $m=n$, $PLoad = \frac{{n+1}^n - 1}{n} \approx {n+1}^{n-1}$. Hence, we achieved almost optimal load balancing. The same ideas hold for $m = n^2, n^3, ..., n^n$ moving one digit to the right in each case.

## 5.2. Key Assignment Scheme B

The second key assignment scheme is more flexible as it will be clear from our discussion. Let $F = \{1, 2, ..., n\}$ be the set of symbols which form a permutation and $S$ be the set of all possible permutations. The symbols in $F$ are common numbers and so we can order them in the ordinary way $1 < 2 < .... < n$. This order implies that we can order $S$ as well according to the lexicographic order. As a result we can present $S$ with its elements ordered: $S = \{Y_1, Y_2, ..., Y_{n!}\}$, $Y_1 <_p Y_2 <_p ... <_p Y_{n!}$. The symbol $<_p$ in an expression as $x <_p y$ obviously denotes that $x$ precedes $y$ in the order of permutations. Let $W = \{1, 2, ..., n!\}$. Our goal is to find a bijection $f : S \to W$ where $f$ will be the mapping function. F plays the role of $value(\cdot)$ mentioned in scheme A.

If we carefully observe the representation of a permutation we can deduce that its first digit from the left can be used to divide $S$ in $n$ classes of permutations with each class containing permutations that have the same leftmost digit. Each class contains $(n-1)!$ members. Thinking in a similar way we can divide further each class to subclasses according to the next digit from left to right stopping at a class with size 1. Each class is equally divided each time we move from one digit to the other. Since permutations are ordered, classes can be ordered in the same manner. Easily then we can find the position of a permutation $X$ in $S$ by counting the total number of permutations belonging to classes preceding $X$. It is straightforward to see that $f$ should be:

$f(x_1, x_2, ..., x_n) =$
$= w_1(x_1)(n-1)! + w_2(x_2)(n-2)! + ... + 1 \cdot 1! + 1$

where $X = x_1 x_2 ... x_n \in S$. Let $S_i = [x_i, x_{i+1}, ..., x_n]$ and $p(x_j)$, $i \leq j \leq n$ be the position number of $x_j$ if we sort $S_i$. Then $w_i(x_i) = p(x_i) - 1$. Now suppose that we would like to create the key space for our TRANS-Net which has $n!$ peers. We assign to each peer $K$ keys where $K$ is a constant system parameter. The first peer is assigned the keys [0,K-1], the second peer the keys [K,2K-1] etc. The last peer is assigned the keys [(n!-1)K, n!K-1]. The total key space is [0, n!K-1 ] and bounds $N_{left}, N_{right}$ are set to values 0 and n!K-1 respectively. Searching for a key x in TRANS-Net which uses mapping function B is easy. We first calculate the peer $P$ who is responsible for x, $P = \left\lfloor \frac{x}{K} \right\rfloor + 1$. Then we calculate the weights $w_i(x_i)$, by solving the following equation $f(x_1, x_2, ..., x_n) = P$, which is a simple task and is accomplished by simple division. Next we calculate $p(x_i)$ and then $x_i$. At last we have the P's permutation id which allows to route through the TRANS-Net. The mapping function B is more flexible than A since it is not restrictive and key space can be customized easily. On the other hand function A is presented in this paper because we believe it is a useful candidate for future P2P design involving networks based on permutations.

## 6. Multi-Dimensional Queries

As we will show shortly our system is capable of handling multi-dimensional queries. Firstly we think of data as a set of n-dimensional arrays of the form: $x = \{x_1,...x_n\}$ where n is our system's parameter and $x_i \in D_1, D_2$, $i=1,...,n$, $D_1, D_2 \in \mathbb{Q}$. Each data array x is normalized (min-max) to an array y, $y = \{y_1,...y_n\}$, for which it holds: $y_i = \left\lfloor \frac{x_i - D_i}{D_2 - D_1} \cdot (n-1) \right\rfloor + 1$. Each data array y is assigned to the node which owns the closest permutation identifier $P = [z_1,...,z_n]$ such as $f(P) \geq f(y)$, $f(\cdot)$ refers to the mapping function used by schemes A and B. What we have accomplished here is to distribute multidimensional data on our network nodes in a meaningful way allowing queries on the data to be answered effectively. We underline here that it is easy to see that range multi-dimensional queries can also be supported by visiting consecutive nodes. We leave as future work immediate applications such as P2P image similarity. One possible restriction of our scheme is the parameter n which is defined as a constant thus determining the maximum number of dimensions.

## 7. Fault-Tolerance and Congestion

In this section we investigate the fault-tolerance and congestion parameters of our system. Bisection width (BW) is a common metric for both the capacity of the network and the fault-tolerance. Large capacity means large tolerance on congestion and failures. Bisection width is of primary importance in the case of high loads with random destinations. Formally bisection width is defined as the smallest number of edges between two partitions of equal size of the graph describing the network under consideration. In [7] a comparison is made of several popular topologies based on their bisection width. For Baton* a newly developed system not mentioned in [7], it is easy to prove its bisection width value, which is Θ(m). From [7],[13] we can see that the full transposition network (regarding the complete topology graph of networks being compared) has a larger value than the value of a star network which is similar to a pancake network. The ratio is $\frac{n}{4}$. Furthermore, we can deduce that for Butterfly networks and De Bruizn networks which have comparable bisection width values, their BW values are larger than the BW of a full transposition network. The ratio for the complete graphs of each topology is Θ(n). CAN and Chord are comparable to Baton* [7]. Vertex connectivity is another fault tolerance metric of a network. Regarding the vertex connectivity metric the full transposition topology is optimal [6]. There are $\frac{n(n-1)}{2}$ distinct paths between two nodes in the network and this is equal to the number of distinct paths for a complete De Bruizn topology of the same number of nodes. In [4] it is proven that a network in order to stay connected when all nodes fail with probability ½ some nodes need to have degree $\Omega(\log m)$, where m is the number of nodes in the system. It is easy to notice from the experimental discussion of the routing table size that as our network gets dense it satisfies this requirement. This can not be fulfilled in all cases by other transposition networks such as pancake and star networks which have a maximum degree of n-1. For example for a complete network of n! nodes a pancake network has a degree n-1 which is far less than logn!. Formally we can prove by induction on n, for n>3, it holds that $\deg ree(pancakeNet) = n-1 < \log n!$. On the other hand, for a complete transposition network we can prove also by induction on n that: $\deg ree(FTN) = \frac{n^2 - n}{2} > \log n!,\ n>2$.

## 8. Implementation

In order to evaluate our system in practice we implemented a simulator and conducted extensive experiments. Both mapping functions described above were fully implemented. The simulation system was implemented in Java and it provides a graphical user interface for basic operations such as network initialization, network parameter setting and statistical metric reporting.

## 9. Experimental Evaluation

In the following paragraphs we present experimental results that are indicative of our system efficiency and give support to the theoretical analysis where it is available. Experiments run on an Intel Core Duo 1.99 GHz computer with 1 Gbs of RAM. All experiments were repeated 40 times to avoid the effect of randomness.

## 9.1. Key Look-up Performance

In the next figures it is shown the average lookup cost in number of hops for the search of a randomly generated key and with the search operation initiated from a random node in a TRANS-Net network of N=100, 500, 1000, 5000, 10000 nodes. In each experiment we chose 100 random nodes among the available ones and executed 100 random search key lookups from each node. Each experiment was executed several times. The parameter *n* was set to values 7, 10. The results were acquired using the greedy algorithm (fig. 3, 4) and the heuristic algorithm (fig. 5, 6). We also compared our P2P system's performance with the one exhibited by Chord.

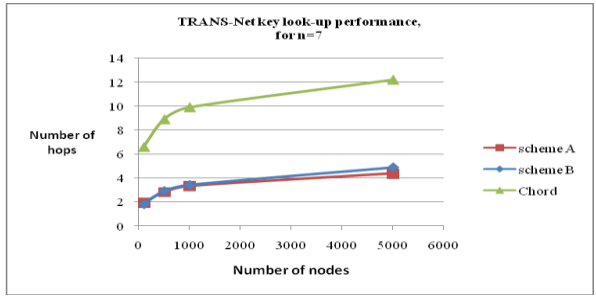

**Figure 3.** Search key look up performance for TRANS-Net for n=7 when greedy algorithm was used.

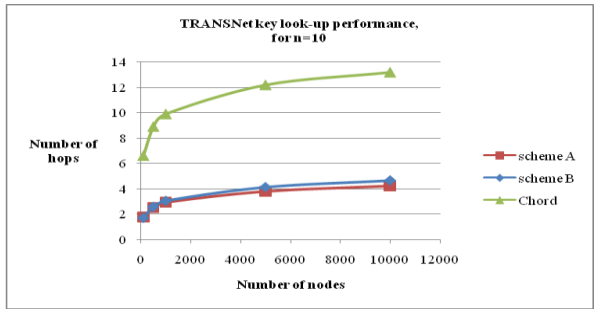

**Figure 4.** Search key look up performance for TRANS-Net for n=10 when greedy algorithm was used.

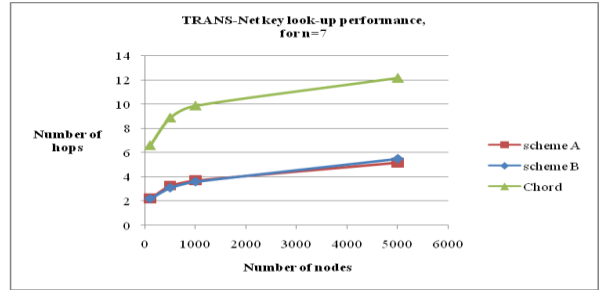

*Figure 5. Search key look up performance for TRANS-Net for n=7 when heuristic algorithm was used.*

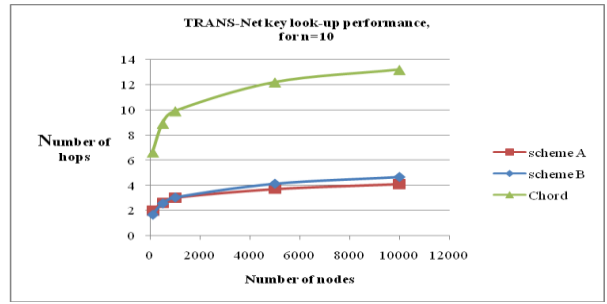

*Figure 6. Search key look up performance for TRANS-Net for n=10 when heuristic algorithm was used.*

The results above for the heuristic algorithm follow our theoretical analysis and they are also comparable to those of the greedy algorithm. We must note here that the mean length path for the heuristic algorithm is counted for successful searches only. Our heuristic algorithm made at most three attempts to find a key restarting a search operation from a different random node found in its routing table when the search operation initially failed. In addition for the key assignment *B*, the *K* parameter doesn't affect the mean path length. It is clear from the results that TRANS-Net shows significantly better key look-up performance than Chord. Compared to other more efficient P2P systems, TRANS-NET exhibits similar performance. For example it is comparable to that of Baton*. For the latest we can tune its performance by adjusting the tree fan out. By tuning n parameter in our system we can also manage performance speed-up. Our heuristic algorithm shows high success rates as it is depicted from the following results (fig. 6, 7):

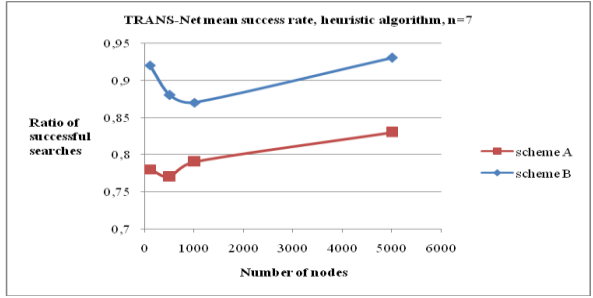

*Figure 7. Mean success rate for the heuristic algorithm, n=7.*

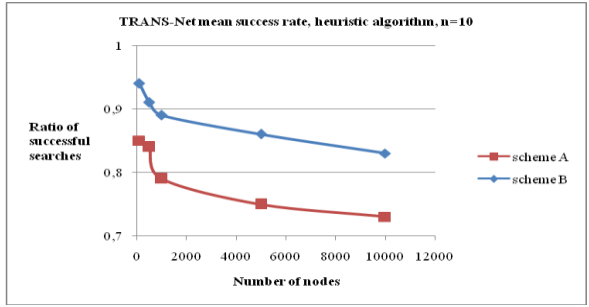

*Figure 8. Mean success rate for the heuristic algorithm, n=10.*

## 9.2. Routing Space Overhead

In this section we investigate the growth in size of the routing table of a node, as the system's number of nodes increases. Specifically, we count the average number of distinct entries in the routing table of a node as the network grows in size.

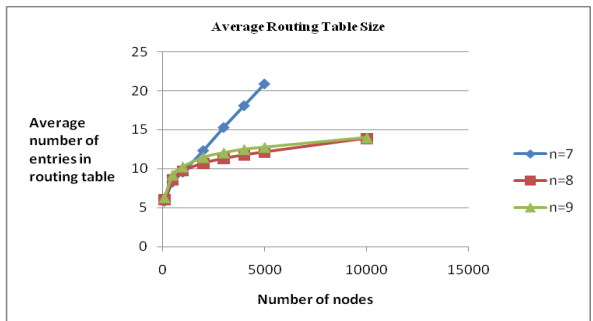

*Figure 9- Average routing table for n=7,8,9.*

From the figures above we can deduce that there is an analogy to the population of the network and the size of the routing table. For example if the network contains half of its maximum number of nodes then it is expected for a node to contain approximately half of the maximum possible distinct entries in its routing table. It is well-known that there is a trade-off between diameter and degree based on the Moore bound and is easy to see that TRANS-NET is not optimal. Generally this is not a problem and our system is not inefficient. For example Baton* has a degree of $(t-1)\log_t m$ where m is the number of nodes in the system. Suppose t=n and m=n!, then we have $(t-1)\log_t m = (n-1)\log_n n!$.

For TRANS-NET we have a degree of $n(n-1)/2$ which is smaller than the degree of Baton*.

## 9.3. Load Balancing Schemes Evaluation

Here we evaluate the two key distribution schemes that we proposed earlier. We verified the theoretical analysis given above by means of extensive experiments. For each scheme, scheme A and scheme B, we conducted an experiment for each value of n=5, 6, 7, 8, 9, 10, 11 and for each value of generated permutations k=5, 10, 100, 500, 1000, 1000, 10000. Each experiment was repeated several times. The experiment was as follows. Firstly we generated k permutations, for some n and sorted them according to their values. Then for each pair of successive permutations $k_i, k_{i-1}$ we calculated the quantity $A = \frac{|PLoad - (value(k_i) - value(k_{i-1}))|}{PLoad}$ where PLoad is the load for each node if keys were evenly distributed, namely $\frac{n!}{k}$. In the next figures we show the mean value of A for the values of n=5, 7, 10, k=5, 10, 100, 500, 1000, 1000, 10000 where k<=n!.

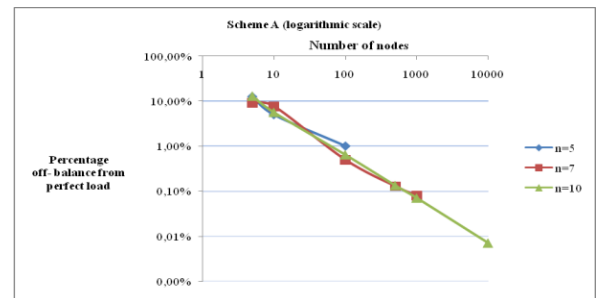

*Figure 10- Mean value of A for scheme A, logarithmic scale for both axes was used with base 10.*

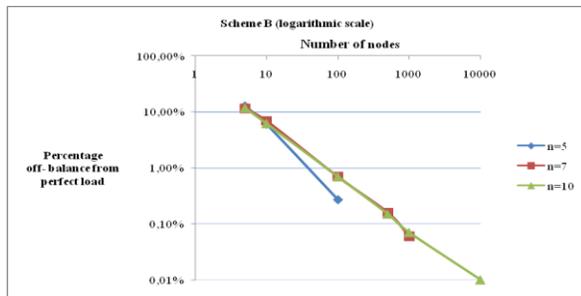

*Figure 11- Mean value of A for scheme B, logarithmic scale for both axes was used with base 10.*

From the results above we can conclude that the schemes distribute evenly the load to the peers of the proposed overlay network and this distribution rapidly approaches the optimal as new peers join the network.

## 10. Conclusion and Future Work

We have just described a highly practical P2P overlay network, TRANS-Net, based on the full transposition network. As far as we know this is the first P2P system designed based on this type of network. Our P2P system efficiently supports multi-dimensional queries. It also exhibits almost optimal load balancing properties on the average, when scheme A is used. Future work includes, but is not limited to, extensive implementation, further more detailed analysis of TRANS-Net and exploitation of the framework for P2P applications such as P2P image similarity.